\documentclass{emulateapj}
\lefthead{GAUDI \& HAIMAN} \righthead{MICROLENSING OF ELLIPTICAL SOURCES}

\def\eq#1{equation (\ref{#1})}
\def\Eq#1{Eq.~\ref{#1}}
\def\vu{{\bf u}}
\def\thetae{\theta_{\rm E}}
\def\dos{D_{\rm os}}
\def\dls{D_{\rm ls}}
\def\dol{D_{\rm ol}}
\def\rsch{R_{\rm S}}

\begin{document}

\title{Microlensing of Elliptical Sources by Fold Caustics}
\author 
{B.\ Scott Gaudi\altaffilmark{1} and Zoltan Haiman\altaffilmark{2}}
\altaffiltext{1}{Harvard-Smithsonian Center for Astrophysics, 60 Garden St., Cambridge, MA 02138}
\altaffiltext{2}{Department of Astronomy, Columbia University, 550 West 120th Street, New York, NY 10027}

\email{sgaudi@cfa.harvard.edu, zoltan@astro.columbia.edu}

\begin{abstract}
~~We consider the problem of an elliptical background source crossing
a linear gravitational lensing fold caustic.  We derive a simple
expression for the light curve of a source with a uniform
surface brightness that is accurate to third
order in the ellipticity $e$ (yielding brightness errors of $\la1\%$
for $e\la 0.3$).  We then consider caustic crossings of a rotating
star, an oblate giant extrasolar planet, and an inclined standard thin
accretion disk around  a quasar black hole, and find the
following results.
(1) If most stars have rotation periods similar to
chromospherically-active (i.e.\ spotted) giant stars in the bulge,
than $\sim 15\%$ should be sufficiently oblate ($e\ga 0.25$) to produce detectable
($\ga 1\%$) deviations with current observations.  The form of the deviation due
to ellipticity is qualitatively similar to that due to limb-darkening.
Thus stellar oblateness, in general, may have to be taken into account in
interpretations of precise limb-darkening measurements with microlensing.  
(2) Giant planets will generally not produce a detectable oblateness
signal, either because they rotate too slowly (if they are close-in
and tidally locked to their parent star), or because they are too
faint for detection (if they are farther away from the parent star).
However, a close-in planet with a large ellipticity could cause
a detectable distortion, especially if the caustic crossing occurs
in the direction parallel to the major or minor axis of the planet's surface.
(3) There is a near-degeneracy between ellipticity and position angle
for equal-area sources with scale-free intensity profiles and
elliptical isophotes.  This degeneracy results in a factor of $\sim 2$
uncertainty in the measurement of the scale-length of a standard thin
accretion disk using observations of a single, linear fold
caustic-crossing quasar microlensing event in passbands probing the
outer parts of the disk.  More precise or higher-frequency
observations can reduce this uncertainty.

\end{abstract}
 
\section{Introduction
\label{sec:intro}}

Gravitational lens caustics have proven enormously useful in
studies of a broad range of astrophysical phenomena.  Caustics are
closed curves that describe the set of source positions where the
Jacobian of the mapping from source plane to image plane induced by
the lens vanishes.  The magnification of a point source on a caustic
in formally infinite.  Caustics are composed of two different
types of singularities: folds and cusps.  Folds are smooth, curved
lines that meet at cusp points.  Cusps are higher-order
singularities where the tangent to the caustic curve is undefined.
The majority of the length of a typical caustic curve is
well-described by a simple fold singularity, and therefore fold
caustics are generally more common.  A source sufficiently close to a
fold caustic is lensed into a pair of equal-magnification,
opposite-parity images whose magnification diverges as $(\Delta
u_\perp)^{-1/2}$, where $\Delta u_\perp$ is the perpendicular
separation of the source from the caustic.  These are the two images
that are created {\it locally} by the fold caustic.  There will generally
be other images created by the global potential of the lens that are
not associated with the caustic.  However, these images will typically
behave in a smooth and continuous manner near the the caustic.  Here we
consider the case of microlensing of sources near fold caustics.  In
this case, the individual images are unresolved, and only the total
magnification of all the images is observable.  Generally, microlensed
sources move on relatively short time scales, and therefore the change
in magnification as a function of time, i.e.\ a light curve, is
observable.  Due to the divergence in magnification as a source
crosses a fold caustic, the light curve can be used to study sources
at higher spatial and angular resolution, than would be possible with
conventional techniques.

Caustics have been proposed or implemented to study objects ranging in
scale from planets to quasars.  Applications relevant to the
discussion here include the direct detection and characterization of
close-in extrasolar giant planets and associated structures via their
reflected light \citep{gg00,li00,al01,gch03}, the precise measurement
of stellar limb-darkening \citep{albrow99,albrow01,fields03,abe03},
and the resolution of the central engines of quasars (e.g.\
\citealt{gkr88,gks91,gm97,ak99,gamm03}).  
We note that, for systems composed of point masses, extended caustics
occur only for composite lensing bodies, the simplest example being a
binary star.  A single star also produces formally divergent
magnification, at the point at which the observer, lens, and source
are perfectly aligned.  Such fortuitously aligned systems can then be
used to resolve stellar surfaces. Indeed, microlensing by a point mass
has been proposed as a tool to study limb darkening
\citep{vallsgabaud98,hendry98,gg99,h03}, stellar spots
\citep{hs00,hendry02}, and ellipticity \citep{hl97}.  The lens and
source must be aligned to better than the angular size of the star for
the source to be resolved in single-lens events.  However, the source
size for typical Galactic microlensing events is less than a few
percent of the angular Einstein ring radius of the lens.  Therefore,
caustic-crossing binary-microlensing events are more common than
source transit events by single stars.

Analytic results for the magnification near fold caustics have been
derived only for circularly-symmetric sources, and only for a few
special functional forms for the surface brightness profile of such
sources (see, e.g.\ \citealt{sef92,albrow99,dominik03}).
However, for a number of applications, one is interested in more
complex source geometries.  Here we consider one such geometry that
should occur frequently in nature: elliptical sources.  In
\S\ref{sec:ellip} we derive a simple, analytic expression for the
change in magnification relative to a circular source that is accurate
to third order in the ellipticity. 
We apply our results in \S\ref{sec:app}, considering the measurement
of stellar oblateness and limb-darkening using fold caustic crossings
(\S\ref{sec:oblate}), the possibility of measuring the oblateness of
giant extrasolar planets (\S\ref{sec:planets}), and the resolution of
a quasar accretion disk (\S\ref{sec:disks}).  We summarize and
conclude in \S\ref{sec:summary}.

\section{Elliptical Sources Near Fold Caustics}\label{sec:ellip}

The total magnification of a point source with an angular separation of $\Delta u_\perp$ normal to a simple linear fold caustic is \citep{sef92},
\begin{equation} 
A(\Delta u_\perp)=\left(\frac{\Delta u_\perp}{u_r}\right)^{-1/2}H(\Delta u_\perp),
\label{eqn:magp}
\end{equation}
where $u_r$ is related to the derivatives of the lens mapping at the fold (see \citealt{sef92,plw01,gp02}),
and $H(x)$ is the Heaviside step function. 
Here and throughout, all angular separations on the sky will be in units of the angular Einstein ring radius of the lens, $\thetae=(2\rsch/D)^{1/2}$, where $\rsch=2GM/c^2$ is the Schwarzschild radius of mass $M$, $D=\dol\dos/\dls$, and $\dos$, $\dos$, and $\dls$ are the distances
between the observer-lens, observer-source, and lens-source.  For cosmological objects, angular diameter distances should be used.  

The magnification of a finite source near a fold is just the convolution of 
\eq{eqn:magp} over the source,
\begin{equation}
A_{\rm fs}=
{{\int_D {\rm d}\vu I(\vu) A(\vu)}
\over
{\int_D {\rm d}\vu I(\vu)}},
\label{eqn:fsgen}
\end{equation}
where $I(\vu)$ is the surface brightness distribution of the source, and $D$ is the solid angle
extended by the source.

\begin{figure}
\epsscale{1.0}
\plotone{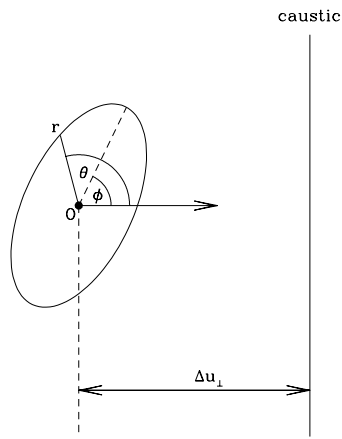}
\caption{\label{fig:one}
The geometry of the integration of the magnification of an 
elliptical source near a fold caustic in a polar ($r,\theta$) 
coordinate system. The origin
is chosen to be the center of the ellipse.  The
major axis of the ellipse is indicated by the dashed line, which is
at an angle $\phi$ with respect to the perpendicular to the caustic.  
}\end{figure}

For the simple case of a uniform--surface--brightness source, adopting a polar coordinate 
system with origin at the center of the source (see Figure \ref{fig:one}), 
\eq{eqn:fsgen} can be
rewritten as
\begin{equation}
A(\eta;e,\phi)=\left(\frac{u_r}{\rho}\right)^{1/2}\frac{1}{\pi}
\int_0^{2\pi}{\rm d}\theta
\int_0^{r(\theta;e,\phi)/\rho} {\rm d}x 
$$
$$\times
\frac{x}{[\eta + x\cos\theta]^{1/2}}
H(\eta+x\cos\theta).
\label{eqn:afsmid}
\end{equation}
Here, $\rho$ is the characteristic angular size of the source, defined
such that the total area extended by the source is $\pi\rho^2$
(i.e. $\rho=r$ for a circle, and $\rho=\sqrt{ab}$ for an ellipse);
$r(\theta;e,\phi)$ is the outer boundary of the source.  We have also
defined the dimensionless angle $x\equiv u/\rho$, and distance from
the caustic, $\eta\equiv\Delta u_\perp/\rho$.  The inner integral in
\eq{eqn:afsmid}
can be evaluated analytically, although care must be
taken to establish the limits of integration so that the condition
$\eta+x\cos{\theta}>0$ is met.

For a circular source, $r=\rho$, and the magnification has the well-known form
(see, e.g.\ \citealt{sef92}),
\begin{equation}
A(\eta,0)=\left(\frac{u_r}{\rho}\right)^{1/2} G_0(\eta),
\label{eqn:a0}
\end{equation}
where
\begin{equation}
G_n(\eta)\equiv \pi^{-1/2} { (n+1)! \over (n + 1/2)!}
\int_{{\rm max}(-\eta,-1)}^{1} {\rm d}x 
$$
$$\times
{(1-x^2)^{n+1/2} \over (x+\eta)^{1/2}} H(1+\eta).
\label{eqn:gfuncn}
\end{equation}
The function $G_0(\eta)$ is shown in Figure \ref{fig:one}.

We now consider a uniform source with arbitrary ellipticity $e$.  
We first address the case of $\eta> 1$, i.e. prior to the source's
entry onto the caustic, for which 
the limits of integration of the inner integral in 
\eq{eqn:afsmid}
are $(0,r/\rho)$ and the limits of the outer integral are $(0,2\pi)$.   
The magnification is then,
\begin{equation}\label{eqn:rinteval}
A(\eta;e,\phi)=\left(\frac{u_r}{\rho}\right)^{1/2}\frac{1}{\pi}
\int_0^{2\pi}{\rm d}\theta\frac{2}{3\cos^2{\theta}} 
$$
$$
\times\{[(r/\rho)\cos\theta-2\eta][(r/\rho)\cos{\theta}+\eta]^{1/2} +2\eta^{3/2}\}.
\end{equation}
For an ellipse with semi-major axis $a$, 
$r=a[\cos^2{\alpha}+(1-e^2)^{-1}\sin^2{\alpha}]^{-1/2}$,
where $\alpha\equiv \theta-\phi$, and $\phi$ is the angle of the major axis
with respect to the normal to the caustic (see Figure \ref{fig:one} for an illustration).
Note that we have centered the
coordinate system at the center (rather than at the focus) of the ellipse.
An ellipse with area $\pi\rho^2$ has $a=\rho(1-e^2)^{-1/4}$, and thus we have,
\begin{eqnarray}
\frac{r}{\rho}&=&[\cos^2{\alpha}+(1-e^2)^{-1}\sin^2{\alpha}]^{-1/2}(1-e^2)^{-1/4},\cr
 &=&1+\frac{1}{4}e^2 \cos2\alpha +O(e^4),
\label{eqn:bndexact}
\end{eqnarray}
Inserting \eq{eqn:bndexact} into \eq{eqn:rinteval}, keeping terms only 
up to order $e^2$, and discarding terms that are odd under $\theta\rightarrow -\theta$,
we find after some algebra,
\begin{equation}
A(\eta;e,\phi)=A(\eta,0)+\left(\frac{u_r}{\rho}\right)^{1/2}e^2\cos(2\phi)E(\eta),
\label{eqn:aephi}
\end{equation}
where we have defined the function
\begin{equation}
E(\eta)\equiv\frac{1}{4\pi}\int_0^{2\pi} {\rm d}\theta
\frac{\cos{2\theta}}{(\cos{\theta}+\eta)^{1/2}}, \hspace{0.5cm} ({\rm for}\,\,\eta>1) 
\label{eqn:ffunc0}
\end{equation}
which can be expressed in terms of elliptic integrals. 

\begin{figure}
\epsscale{1.0}
\plotone{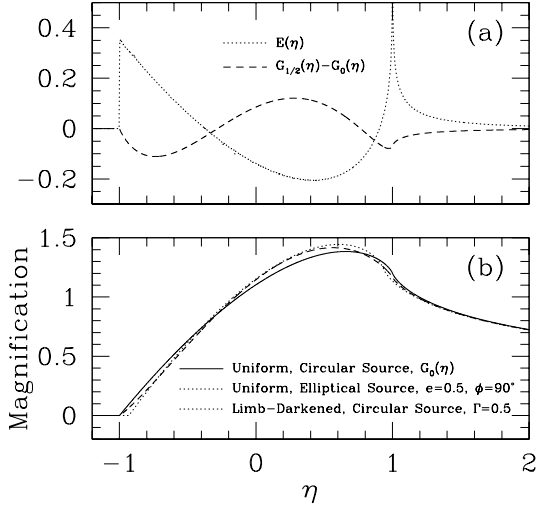}
\caption{\label{fig:two}
(a) The dotted line shows the function $E(\eta)$, the analytical
approximation describing the normalized
difference in magnification between an equal-area elliptical and
circular uniform source as a function of the angular separation $\eta$
between the center of the source and the caustic in units of the
angular size of the source $\rho$. 
For sources of ellipticity
$e$ and position angle $\phi$, this curves should be multiplied
by $e^2\cos(2\phi)$.  See equation~\ref{eqn:aephi}. 
The dashed line shows
$G_{1/2}(\eta)-G_{0}(\eta)$, which is the function describing the
difference in magnification between a uniform and limb-darkened
circular
source.   (b) The solid line shows the normalized light curve
$G_0(\eta)$ for a uniform, circular source crossing a fold caustic.
The dotted line shows the exact light curve
for a uniform elliptical source with ellipticity $e=0.5$ and
$\phi=90^\circ$.  The dashed line shows
$G_0+\Gamma[G_{1/2}(\eta)-G_0(\eta)]$, the light curve for a circular,
limb-darkened source with limb-darkening parameter $\Gamma=0.5$.  For
all curves in both panels, the magnification for a source of size $\rho$ can be
found by multiplying by $\rho^{-1/2}$.
}\end{figure}

\begin{figure}
\epsscale{1.0}
\plotone{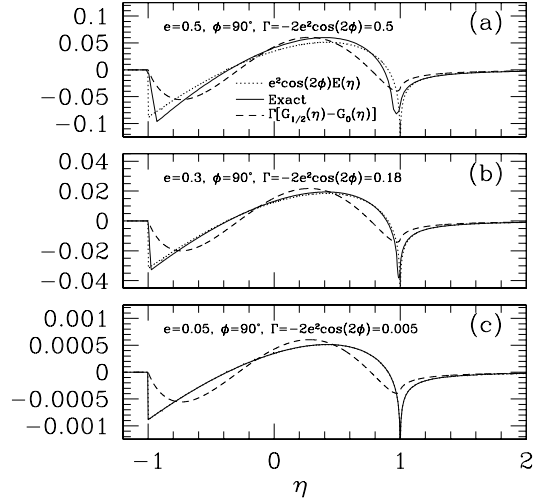}
\caption{\label{fig:three}
In all panels, the dotted line shows the function 
$e^2\cos(2\phi)E(\eta)$, the analytical
approximation describing the difference in magnification between an 
elliptical source with ellipticity $e$ and position angle $\phi$,
and an equal-area, circular, uniform source,  as a function of the angular separation $\eta$
between the center of the source and the caustic in units of the
angular size of the source $\rho$. 
The solid line shows the exact solution for the same
quantity, obtained numerically from equation~\ref{eqn:afsmid}.
The dashed line shows $\Gamma[G_{1/2}(\eta)-G_{0}(\eta)]$, 
the difference in magnification between a circular, limb-darkened source
with limb-darkening parameter $\Gamma$,
and a circular, uniform source. Here $\Gamma=-2e^2\cos(2\phi)$, which approximately reproduces the deviation due to ellipticity. For
all curves, the magnification for a source of size $\rho$ can be
found by multiplying by $\rho^{-1/2}$.
 (a) $e=0.5$, $\phi=90^\circ$, and $\Gamma=0.5$.
(b) $e=0.3$, $\phi=90^\circ$, and $\Gamma=0.18$.
(c) $e=0.05$, $\phi=90^\circ$, and $\Gamma=0.005$.
For ellipticities as small as shown in panel (c), the analytical approximation becomes near-exact, and the dotted and solid
curves are indistinguishable.
}\end{figure}

These results assume that the source is off the caustic.  During
caustic crossing, when $|\eta|\leq 1$, the limits of integration become
complicated.  However, in the limit of $e\rightarrow 0$, it is easy to
see that the last equation would be modified as
\begin{equation}
E(\eta)\equiv\frac{1}{4\pi}\int_0^{2\pi} {\rm d}\theta
\frac{\cos{2\theta}}{(\cos{\theta}+\eta)^{1/2}} H(\cos\theta+\eta).  
\hspace{1cm} 
$$
$$
({\rm for}\,\,   \left|(|\eta|-1)\right|\ga e^2/4)
\label{eqn:ffunc}
\end{equation}
This function is shown in Figure \ref{fig:two}.
In fact, equation (\ref{eqn:aephi}), together with (\ref{eqn:ffunc}),
is valid to third order in $e$ for nearly all values of $\eta$.  
This is because, for small $e$, the appropriate limits of integration of
\eq{eqn:afsmid} are approximately equal for an elliptical and circular source.  
The small changes to the integration limits introduce $O(e^2)$  corrections 
to a quantity that already includes $O(e^2)$ terms; the accuracy of this 
statement can also be verified directly in the numerical results displayed 
in Figure~\ref{fig:three} below.
Since changes to the integration limits are small, the difference in 
magnification between an elliptical and circular source
can be calculated simply by taking the difference between
the radius of an ellipse and circle, $\Delta r=(r-\rho)$, times the
area element $r{\rm d}r{\rm d}\theta$, weighted by the local
magnification, and integrated over all angles such that
$\cos{\theta}+\eta >0$.  However, this approximation breaks down for
$|\eta|\sim 1$, where the limb of the source is near the caustic.
First consider sources just exiting the caustic with $\eta \sim -1$.
For $0\le \phi\le 45^\circ$, the limb of a circular source will exit
the caustic first, whereas part of the limb of an elliptical source
will still be inside the caustic.  The situation is reversed for
$45^\circ\le \phi\le 90^\circ$, when the elliptical source will exit
the caustic first.  This effect is not accounted for in \eq{eqn:ffunc}.
This formalism also breaks down near $\eta = 1$ for similar reasons.
As a result, \eq{eqn:ffunc} cannot be used when 
$\left|(|\eta|-1)\right|\la e^2/4$.  

Since there are no terms containing odd powers of $e$, corrections to
\eq{eqn:aephi} are of order $e^4$.  Therefore, the approximation is
very accurate for rather high values of the ellipticity.  We find that
it predicts the magnification to better than $1\%$ for $e\le 0.3$ and
$\left|(|\eta|-1)\right|\ge e^2/4$.
Figure \ref{fig:three} compares the difference in magnification
given by equations (\ref{eqn:aephi}) and (\ref{eqn:ffunc}),
with the exact calculation using \eq{eqn:afsmid} for three
different values of the ellipticity, $e=0.5, 0.3$, and $0.05$. 

\begin{figure}
\epsscale{1.0}
\plotone{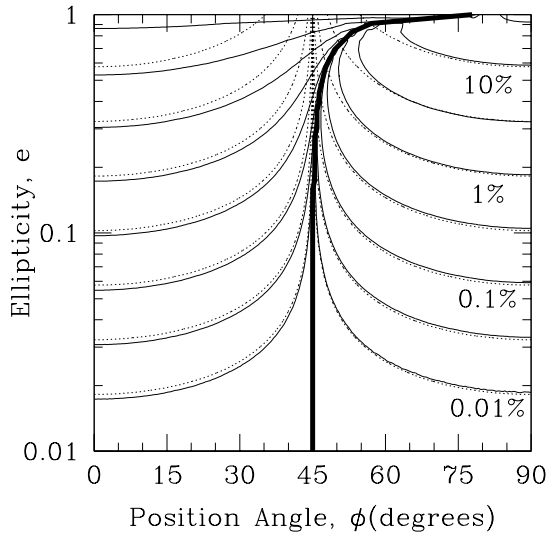}
\caption{\label{fig:four}
Contours of constant RMS fractional deviation of the light curve of a uniform,
elliptical source from an equal-area uniform circular source, as a
function of the ellipticity $e$ and position angle $\phi$ of the
source.  We have assumed that 20\% of the light 
is contributed by sources unrelated to the fold
caustic images.  The solid contours are exact, the dotted contours are the
approximation $0.3 e^2\cos{2\phi}$.  The heavy solid curve shows the
locus of $(e,\phi)$ where the deviation due to ellipticity vanishes.
Contours are at equally-spaced logarithmic intervals of 0.5 dex.
}\end{figure}

So far, we have parameterized the magnification in terms of $\eta$,
the perpendicular angular separation of the center of the source from
the caustic. To convert to magnification as a function of 
time $t$, we can replace $\eta \rightarrow (t-t_0)/\Delta t$, 
where $t_0$ is the
time when the center of the source crosses the caustic, and we have
defined $\Delta t \equiv \rho t_E \csc\gamma$.  Here $t_E$ is the time
is takes the source to cross an angle $\theta_E$, and $\gamma$ is the
angle the trajectory of the center of the source makes with respect to
the caustic.  One cannot constrain $t_E$ and $\gamma$ separately from a
measurement of $\Delta t$ using observations of a single linear fold
caustic crossing, since $t_E$ is degenerate with $\gamma$.  
However, with two crossings,
which will generally have two different values of $\gamma$, both
parameters can be measured.

In terms of detectability, the quantity of interest is
root-mean-square (RMS) fractional deviation $\sigma$ between an
elliptical and circular source.  Under the approximation
\eq{eqn:aephi}, this is simply
\begin{equation}
\sigma(e,\phi) = Ae^2 \cos{2\phi}, 
\label{eqn:rms}
\end{equation}
where 
$A \equiv (\eta_2-\eta_1)^{-1} \int_{\eta_1}^{\eta_2} {\rm d}\eta
E(\eta)/(G_0(\eta)+0.2)$ is the RMS of the normalized fractional
deviation due to ellipticity over the range $(\eta_1,\eta_2)$.  The
total magnification of images not associated with the caustic is
generally of order unity.  For typical source sizes of $\rho \sim
0.05$, the mean total magnification of the fold images during the
crossing ($|\eta| \le 1$) is $\sim \rho^{-1/2} \sim 5$.  We have
therefore assumed a $20\%$ contribution from images unrelated to the
fold caustic.  For $-1\le \eta \le 1.5$, $A\simeq 0.3$.  We note that
this value of $A$ depends not only on the adopted value of the
contribution from images unrelated to the source, but also on the
time-interval $(\eta_1,\eta_2)$ over which the RMS is evaluated.
Changing this interval will
increase or decrease the value of $A$, and therefore change the 
evaluated RMS fractional deviation.   We 
have chosen the interval $-1\le \eta \le 1.5$,
because this essentially the largest interval where the deviation from ellipticity is
significant (see Figure \ref{fig:two}), and therefore this is the interval
of interest in terms of signal-to-noise. 

 Figure \ref{fig:four} shows the
RMS deviation in this range as a function of $e$ and $\phi$, for both
the approximation in \eq{eqn:rms}, and the exact calculation.  It is
clear the the approximation is excellent for $e\la 0.3$.  It is
interesting to note that the deviation due to ellipticity vanishes for
some value of $\phi$, for all values of $e$.  For $e\ll 1$, this
occurs at $\phi=45^\circ$, whereas this null shifts to larger values
of $\phi$ as $e\rightarrow 1$. Interestingly, for large ellipticities,
the light curve from a circular source is degenerate with that from
an ellipse with its major axis aligned nearly parallel to the caustic.

It is clear from the form of \eq{eqn:aephi} that one cannot measure
the ellipticity of a source from one caustic crossing alone; rather
one measures only the quantity $e^2\cos{2\phi}$.  If the source
crosses multiple caustics, each of which will generally have a
different value of $\phi$, then it will be possible to measure $e$ and
$\phi$ separately.

\section{Applications}
\label{sec:app}

\subsection{Stellar Oblateness and Limb Darkening}\label{sec:oblate}

Microlensing has been proposed and employed as a method of studying
various topics in stellar astrophysics (see, e.g. \citealt{gould01} for a
review).  The fold caustic-crossings created by binary-lens events
toward the Galactic bulge and Magellanic clouds are exceptionally
useful in this regard, as they are relatively
frequent\footnote{Approximately $10\%$ of all microlensing events are
caustic-crossing binary-lens events \citep{alcock2000,udalski2000}, whereas
less than a few percent of microlensing events are single-lens events
in which the source is resolved.  Therefore, the majority of events for which
it is possible to resolve the source are binary events.},
and they can be anticipated \citep{jm01}.

Here we consider whether the oblateness of stars can be measured in
caustic-crossing binary-lens events.  The oblateness $f$ of a star is
defined as the fractional difference between its  equatorial and polar
radii.  This can be related to its rotation rate (e.g.\
\citealt{sh02}),
\begin{equation}
f=\frac{R^3}{2GM}\left(\frac{P}{2\pi}\right)^{2},
\label{eqn:oblateness}
\end{equation}
where $M$, $R$, and $P$ are the mass, radius, and rotation period of
the star, respectively.  The ellipticity of the star is then
$e=[1-(f+1)^{-2}]^{1/2}$.

We first estimate the distribution of ellipticities of giant stars in
the Galactic bulge.  \citet{olech96} has compiled a catalog of
miscellaneous periodic variable stars in the Galactic bulge found by
the OGLE II microlensing survey, which includes the period, de-reddened
color and magnitude of each star.  Most of these stars are red ($V-I
>1$), and are in the giant branch or red clump.  \citet{olech96}
argues that most of these red variables, which comprise $\sim 1\%$ of
all stars in this part of the color-magnitude diagram, are likely to
be chromospherically-active (i.e.\ spotted) stars.  Under the
assumption that all are single stars, and that the measured period
corresponds to the rotation period of the star, we use this
catalog to estimate the distribution of oblateness for such
chromospherically-active stars in the bulge.  Since these stars are
likely bulge giants, they will have masses of $M\sim M_\odot$.  We
estimate their angular radii from their $(V-I)_0$ color and $I_0$ magnitude,
employing the color-surface brightness relation of \citet{vb99}, as
presented in \citet{albrow00}.  We then determine
their physical radii by assuming that they are at a distance
of $8~{\rm kpc}$.   We determine the oblateness and
ellipticity of each star using \eq{eqn:oblateness}.

\begin{figure}
\epsscale{1.0}
\plotone{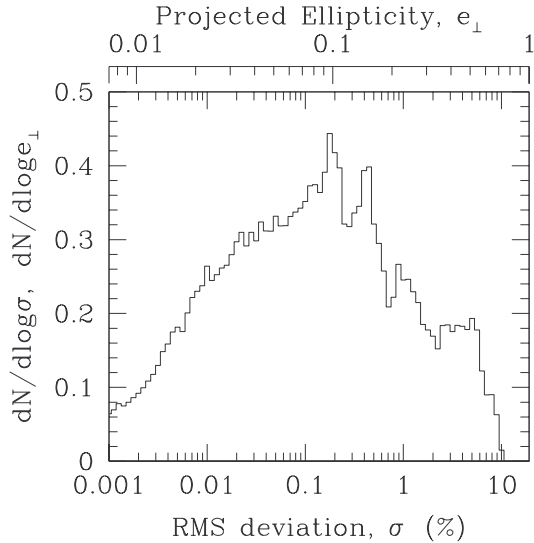}
\caption{\label{fig:five}
The distribution of projected ellipticities of 551 
chromospherically active giant stars in the Galactic bulge, as 
inferred from their 
colors, magnitudes, and rotation periods measured by 
OGLE II \citep{olech96}.  The top axis labels show the inferred
ellipticity, and the bottom axis labels shows the RMS light curve 
deviations that would occur (relative to a circular source) if the 
stars were to cross randomly oriented lensing caustics.
}\end{figure}

The three-dimensional ellipticity of a source is not observable in
caustic-crossing events, rather only the ellipticity projected on the
plane of the sky, $e_\perp = e (1-\cos^2{i})$.  Here $i$ is the angle
between the line-of-sight and the rotation axis of the star.  For a
ensemble of stars, this angle will be distributed uniformly in
$\cos{i}$.  The probability distribution of $e_\perp/e$ is thus,
$P(e_\perp/e)=\frac{1}{2}[1-(e_\perp/e)]^{-1/2}$.  

We note that there is a selection effect that introduces a bias in our
determination of the distribution of projected ellipticities, which we
make no attempt to correct.  Nearly pole-on stars will have smaller
variability amplitudes, and therefore will be preferentially missed by
the OGLE survey.  Therefore, our assumption that the distribution of
$\cos{i}$ is uniform is not actually correct for this sample.  We
expect this bias to be relatively small, because systems have to be
very close to pole-on to be missed (due to the excellent quality
of the OGLE photometry), and because this bias will largely be washed
out by the intrinsic scatter in the amplitudes of the photometric 
modulation due to, e.g.\ variable numbers and distributions of spots.

Figure \ref{fig:five} shows the resulting distribution of projected
ellipticities for the 551 stars from the \citet{olech96} catalog. 
The median ellipticity is $e=0.07$, with $\sim 20\%$ of stars have
$e>0.2$.  We relate the projected ellipticity to the RMS light curve
deviation $\sigma$ by integrating \eq{eqn:rms} over $\phi$, which
yields $\sigma \simeq 0.2 e_\perp^2$.  The median value is $\sigma
\sim 0.1\%$, with $\ga 15\%$ of stars expected to produce RMS
deviations $\ga 1\%$.

Current microlensing follow-up efforts can regularly achieve
single-exposure precisions of $1\%$ on bright giant stars. Therefore,
if the majority of giant stars in the bulge have rotation periods that
are similar to those in the \citet{olech96} sample, then $10-20\%$ of
all caustic-crossing events should show deviations due to the stellar
oblateness.  Whether or not chromospherically quiescent stars 
should have similar rotation periods to active stars is not
clear.  Our analysis demonstrates that this question could be answered with
a sample of $\sim 30$ precise, well-covered caustic-crossing events.

\begin{figure}
\epsscale{1.0}
\plotone{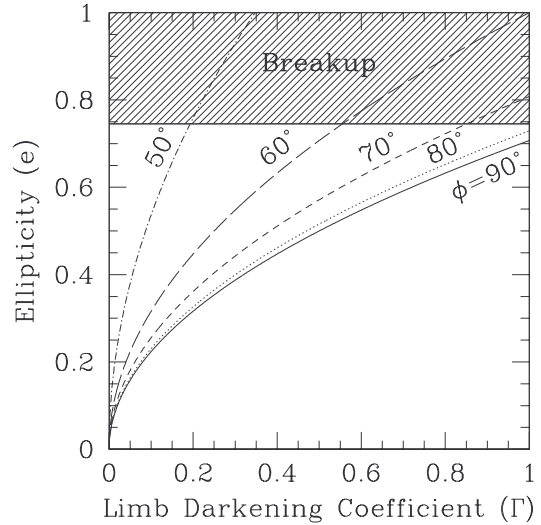}
\caption{\label{fig:six}
The lines show the relation $e=[-0.5\Gamma \sec{2\phi}]^{1/2}$, 
the ellipticity $e$ required for a uniform, elliptical
source to approximately reproduce the light curve of a limb-darkened circular source, as a 
function of the limb-darkening parameter $\Gamma$, for position angles of the
elliptical source relative to the caustic normal of
$\phi=0, 10^\circ, 20^\circ, 30^\circ$, and $40^\circ$.   Sources with $\phi=45^\circ$
produce no deviation due to ellipticity, and sources with $\phi<45^\circ$ produce
deviations that are incompatible with limb-darkening.  Values
of $e\ge 0.745$ are not
allowed for rotating stars, as they would be rotating faster than breakup speed.
}\end{figure}

One potential complication to the measurement of oblateness is limb-darkening.  
For a circular, limb-darkened
source with radial surface brightness profile of the form
\begin{equation}
I(r)=1-\Gamma
\left\{ 1-\frac{3}{2} \left[1-\left(\frac{r}{\rho}\right)^2\right]^{1/2}\right\},
\label{eqn:ldsf}
\end{equation}
the magnification of a limb-darkened source is \citep{albrow99},
\begin{equation}
A(\eta;\Gamma)=A(\eta,0)
+\left(\frac{u_r}{\rho}\right)^{1/2}\Gamma [G_{1/2}(\eta)-G_0(\eta)].
\label{eqn:ald}
\end{equation}
The form for the surface-brightness profile given in \eq{eqn:ldsf} was
introduced in \citep{albrow99}.  It is has the same behavior
as the usual linear limb-darkening parameterization, but has the
advantage that there is no net flux associated with the limb-darkening.
The normalized limb-darkening function $G_{1/2}(\eta)-G_0(\eta)$ is shown
in Figure \ref{fig:two}.

It is clear that the form of the deviation due to the limb-darkening
is qualitatively similar to the deviation due to oblateness for
$\phi>45^\circ$, such that a limb-darkening circular source with
$\Gamma \sim -2 e^2 \cos(2\phi)$ approximately reproduces the light
curve of a uniform elliptical source with parameters $e,\phi$.  
The factor of two is the ratio of the RMS deviations of $E(\eta)$ and
$G_{1/2}(\eta)-G_0(\eta)$ for $-1 \le \eta \le 1$.  The value
of $e$ required to produce a given limb-darkening $\Gamma$ is plotted
in Figure \ref{fig:six}, for several values of $\phi$.  
Figure \ref{fig:two} shows an
example, for a limb-darkened circular source with $\Gamma=0.5$, and a
uniform, elliptical source with $e=0.5$ and $\phi=90^\circ$.  
Other examples are shown in Figure~\ref{fig:three}. The
degeneracy is only approximate, however the two curves agree well
except near the beginning and end of the caustic crossing.  Note that,
in general, the time when the source exits the caustic (near $\eta \sim -1$)
will be different for circular and elliptical sources of equal area.

Of course, stars will, in general, be both oblate and limb-darkened.  
We find numerically that for small ellipticities, the light curve of 
an elliptical, limb-darkened source is well-approximated by the superposition 
of the effects of limb-darkening and ellipticity.  In the optical to
near-infrared, limb-darkening parameters for typical microlensing
stars are in the range $\Gamma=0.3-0.7$ \citep{fields03}, and
therefore limb-darkening effect will typically dominate over the
effect of ellipticity. However, given the extraordinary precision of
recent limb-darkening measurements \citep{fields03,abe03}, it is not
clear that the effect of oblateness can be ignored.  In particular,
the inconsistency claimed by \citet{fields03} between
derived
limb-darkening parameters of the K3 III source of microlensing event
EROS BLG-2000-5 and stellar atmospheric
model model predictions may be partly reconciled by allowing for stellar 
oblateness. 
Fortunately, in this case a spectrum is available from which the
$v\sin{i}$ of the star can be constrained.  However, in the future,
modelers should be aware of this possible contamination to
limb-darkening measurements.

How can oblateness be distinguished from limb-darkening?  First, since
the effects are not completely degenerate, this may be possible to
distinguish between them simply from high-precision, single-color measurements.
Second, the oblateness signal is expected to be achromatic, whereas
limb-darkening depends strong on wavelength.  Finally, one can use
multiple caustic crossings to attempt to distinguish between
limb-darkening and oblateness: a second caustic crossing can be
anticipated, and in general, it would occur at a different angle
$\phi$ from the first.

\subsection{Oblateness of Extrasolar Planets\label{sec:planets}}

\citet{gg00} and \citet{li00} demonstrated that close-in, giant
planetary companions to the source stars of binary microlensing events
can be detected if the planet crosses a fold caustic, which magnifies
the reflected starlight to detectable levels.  Subsequently,
\citet{al01} and \citet{gch03} explored the detectability of
structures associated with planets using this method, such as the
planetary phase, atmospheric features, satellites, and rings.  These
authors found that, while the planet itself may be detectable, as well
as variations due to the planet phase and associated rings, all other
features are likely to be undetectable with foreseeable telescopes.

Can the oblateness of giant planets be detected via this method?  
\citet{gch03} calculated the
expected signal-to-noise ratio $Q_p$ of the primary planet signal
for a typical event. They found $Q_p \sim 150 (D/100{\rm m})^2 (a/0.05~{\rm AU})^{-2}$, where $D$ is the telescope aperture 
and $a$ is the semi-major axis of the planet orbit. From \eq{eqn:rms},
the signal-to-noise ratio $Q_e$ of the deviation 
due to ellipticity, averaged over all $\phi$, can be related to $Q_p$ by 
\begin{equation}
Q_e \sim 0.2 e^2 Q_p \sim 30 e^2 \left(\frac{D}{100{\rm m}}\right)^{2}
\left(\frac{a}{0.05~{\rm AU}}\right)^{-2}.
\label{eqn:snrp}
\end{equation}    
Thus we can expect
signal-to-noise ratios of less than unity for the deviation due to
ellipticity, unless the ellipticity is quite large, $e \ga 0.2$.
On the other hand, close-in giant planets with $a\la 0.2~{\rm AU}$ are
expected to be tidally locked to their parent star, and therefore their
oblateness due to rotation are expected to be quite small, $f\la
0.3\%$ or $e\la 10\%$ \citep{sh02}.
Planets whose rotation periods
have not been synchronized with their orbital periods may have large
oblateness.  For example Jupiter has an oblateness of $f\sim 0.065$,
which corresponds to an eccentricity of $e\sim 0.34$.  Unfortunately,
the amount of reflected light, and therefore the signal-to-noise
ratio, falls off as $a^{-2}$, decreasing the signal-to-noise ratio of
the planet signal by a factor of $\ga 16$ for tidally-unaffected planets
with $a\ga 0.2$, and therefore rendering the
distant planets too faint for the oblateness signal to be detectable, 
even if they are rapidly rotating.

\subsection{Accretion Disks\label{sec:disks}}

Numerous authors have considered the idea of using microlensing of
multiply-imaged quasars to resolve the surface brightness distribution
of the central accretion disk.  Studies range from theoretical
examinations of the feasibility and details of the method itself
(e.g., \citealt{gkr88,gks91,ak99}), to detailed fitting of light
curves of the lens Q2237+0305 (e.g.,
\citealt{yonehara01,shalyapin02,gamm03,kochanek03}).  The majority of
these studies considered source models of face-on accretion disks.
This assumption is generally not appropriate for standard
geometrically-thin accretion disks: disks are more likely to be seen
edge-on than face-on.  Here we briefly consider the signature of
inclined accretion disks on fold caustic-crossing light curves, and in
the process uncover a degeneracy between the inclination and position
angle for disks with self-similar intensity profiles.  For simplicity,
we consider only linear fold caustics. However, we note that this
assumption may be inappropriate for typical quasar microlensing
scenarios, which have relatively large sources, and optical depths to
microlensing near unity \citep{wyithe00,kochanek03}.  Lensing of an
inclined disk by a point mass was considered by \citep{hl97}.

\begin{figure}
\epsscale{1.0}
\plotone{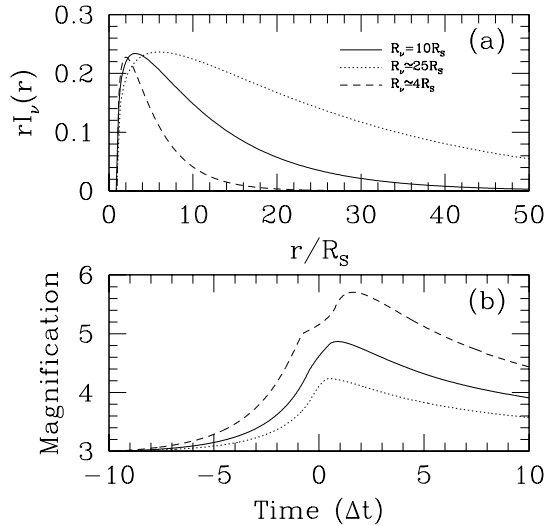}
\caption{\label{fig:seven}
(a) Normalized surface brightness profiles $rI_\nu(r)$ as a function
of radius $r/\rsch$ in units of the  Schwarzschild radius,
for a standard thin accretion disk with inner radius $3\rsch$.  Profiles
are shown for three different values of the radius $R_\nu$ where
the temperature of the accretion disk is equal to $h\nu/k$: $R_\nu=10\rsch$ (solid), $R_\nu\simeq 25\rsch$ (dashed), and $R_\nu \simeq 4 \rsch$ (dotted). 
(b) Magnification as a function of time in units of $\Delta t=R_\nu/v_e$, where $v_e$ is the effective source transverse velocity.  The curves correspond
to the surface brightness profiles in panel (a).  We have assumed that the disk has inclination $i=30^\circ$, position angle relative to the caustic normal of $\phi=15^\circ$, an Einstein ring radius of $\thetae\dos=100\rsch$, and 
that the magnification outside the caustic is $A_0=3$.
}\end{figure}

We consider a standard, optically-thick, geometrically-thin accretion
disk \citep{ss73}.  The disk radiates locally as a blackbody
with temperature,
\begin{equation}
T(r)=T_\nu \left(\frac{r}{R_\nu}\right)^{-3/4}\left[1-\left(\frac{r}{3\rsch}\right)^{-1/2}\right]^{1/4},
\label{eqn:temp}
\end{equation}
where $r$ refers to the radial distance of a (face-on) disk from the 
center, and we have assumed an inner disk edge of $3\rsch$. 
Here $R_\nu$ is the radius at which the local temperature of the disk
matches the wavelength of observations, i.e.\ where $T(r)=T_\nu=h\nu/k$.
Under the assumption of a standard thin-disk model, 
$R_\nu$ can be related to the mass of the black hole and the mass accretion rate.
The surface brightness profile is then,
\begin{equation}
I_\nu(r)\propto \left({\rm e}^{T_\nu/T(r)}-1\right)^{-1}.
\label{eqn:tdsbp}
\end{equation}
Note that we have ignored all relativistic and Doppler effects.
Figure \ref{fig:seven} shows the normalized surface brightness profile
for the assumption of $R_\nu=10\rsch$, which roughly corresponds to
the value found by \citet{kochanek03} from an analysis of OGLE light
curves of Q2237+0305 in the $V$-band, corresponding to rest-frame
$2000$\AA.  We also show surface brightness profiles for
$R_\nu=2^{4/3}10\rsch$ and $R_\nu=2^{-4/3}10\rsch$, i.e.\ for half and
twice the frequency of observations, respectively.

We calculate the light curves by first determining the image areas 
of a finite number of concentric elliptical
annuli using the exact form for the magnification (\Eq{eqn:afsmid}).
The ellipticity of a disk with inclination angle $i$ 
(where $i=0$ is face on) is $e=(1-\cos^2{i})^{1/2}$. 
We weight each annulus by the local surface brightness, which
is simply given by \eq{eqn:tdsbp}, with $r$ replaced by the
semi-major axis of the elliptical annulus.  The magnification is
then the sum over all annuli of the surface-brightness
weighted image area, divided by the surface-brightness weighted
area of the source.  We integrate out to $r=10R_\nu$, beyond which we find that
the contribution to the magnification is negligible. 
Figure \ref{fig:seven} shows light curves for a standard thin accretion disk
with inclination $i=30^\circ$ and position angle relative to the caustic
normal of $\phi=15^\circ$.
The units of time are $\Delta t=R_\nu/v_e$, where $v_e$ is the effective source transverse velocity.
In addition, we have adopted parameters appropriate to
 Q2237+0305, namely an Einstein ring radius of $\thetae\dos=100\rsch$, and
a total magnification of images unrelated to the caustic of $A_0=3$ 
(see, e.g.\ \citealt{kochanek03}).  For this system,
and $V$-band observations, $\Delta t \sim 60~{\rm days}$.

\begin{figure}
\epsscale{1.0}
\plotone{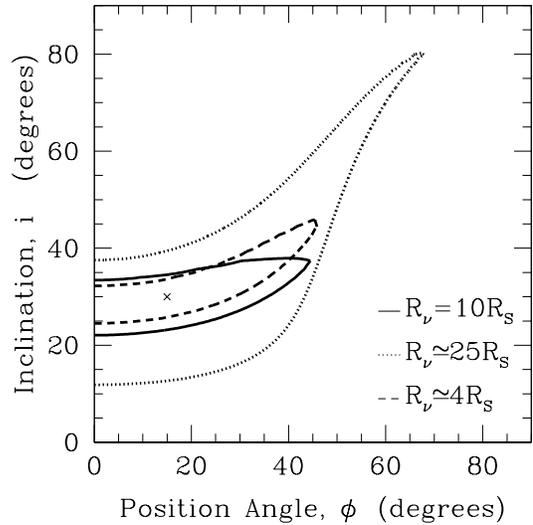}
\caption{\label{fig:eight}
Contours show the $1\sigma$ confidence limits on the inclination $i$
and position angle $\phi$, for simulated observations of a thin accretion
disk crossing a linear fold caustic, with input parameters 
 $i=20^\circ$ and $\phi=15^\circ$, shown as the cross. 
We have assumed 100
measurements from $-10 \le \Delta t \le 10$, each with $2\%$ error,
and a background magnification of $A_0=3$.  
The lines types correspond to light curves in Figure \ref{fig:seven},
and are for $R_\nu=10\rsch$ (solid), $R_\nu\simeq 25\rsch$ (dashed),
and $R\nu \simeq 4 \rsch$ (dotted).
}\end{figure}

In \S\ref{sec:ellip} we demonstrated analytically that 
the light curves of uniform, equal-area elliptical
sources with $e\ll 1$ are degenerate in $e$ and $\phi$ , 
such that one can only measure the combination $e^2\cos{2\phi}$.  
In fact, as could have been anticipated from
Figure \ref{fig:four}, we find numerically that the degeneracy between $e$ and $\phi$ exists for arbitrary ellipticities, again provided that the sources are normalized to have equal area.  In turn, this implies that the degeneracy
between $e$ and $\phi$ exists for any source with surface brightness profile 
that is elliptical and scale-free.  The scale-free requirement arises
from the fact that the ellipses must have equal-area, which implies 
different values for the semi-major axes. Therefore the
introduction of a scale in the surface brightness profile will break
the degeneracy.  For accretion disks, the degeneracy translates into
a degeneracy between $i,\phi$, and $R_\nu$.  However, since the
profile $I_\nu(r)$ of a standard thin disk (\Eq{eqn:tdsbp}) is {\it
not} scale-free, this degeneracy is not perfect.  The severity of the
degeneracy is set by the radius of the inner edge of the disk relative
to $R_\nu$.  If $\rsch/R_\nu \ll 1$, the surface brightness
profile will essentially be scale-free, and the light curves will be
degenerate.  
However, if $\rsch/R_\nu \sim 1$, the central
hole in the surface brightness profile of the disk creates a feature
in the light curve (see Figure \ref{fig:seven}).   Larger inclinations
will produce more pronounced features due to this central hole, 
thus allowing one to distinguish the light curve produced from 
different inclinations.   We note that this degeneracy was
likely present in the simulations of \citet{ak99}, although it does not seem
to have been explicitly recognized as such. 

To crudely quantify the severity of this degeneracy, we have simulated
observations of a caustic-crossing event.  We adopted the parameters
of the light curves shown in Figure \ref{fig:seven}, namely
$i=30^\circ$, $\phi=15^\circ$, $\thetae\dos=100\rsch$, and $A_0=3$.
We assume 100 equally-spaced observations from $-10 \le \Delta t \le
10$, each yielding photometry in a single band with a fractional
flux error of $2\%$. 
For the parameters of Q2237+0305, this
corresponds to a sampling interval of $\sim 12~{\rm days}$.  This
sampling interval and photometric error are comparable to that in the
OGLE light curve of Q2237+0305 \citep{wozniak00}.  Figure
\ref{fig:eight} shows the resulting $1\sigma$ confidence regions in the
$i,\phi$ plane for the three different light curves.  As expected, the
degeneracy is most severe for $R_\nu=25 \rsch$, where inclinations as
large as $80^\circ$ are consistent with the simulated light curve at
the $1\sigma$ level.  This translates into a factor of
$(\cos{30^\circ}/\cos{80^\circ})^{1/2}\sim 2$ uncertainty in $R_\nu$.  Note that
we have fixed all other free parameters at their input values; allowing these
parameters to vary would worsen the degeneracy.

Since a measurement of $R_\nu$ can be used to constrain the mass of
the black hole, it is important to reduce the uncertainty in $R_\nu$
if possible.  This can be done by obtaining higher-precision
observations, although we find that, for the same sampling rate, the
precision of the individual measurements must be $\la 0.2\%$ to reduce
the uncertainty to below $20\%$.  A more robust way of reducing this
uncertainty is to observe at higher frequencies, where the ratio
$\rsch/R_\nu$ is larger.  For example, for $R_\nu=10\rsch$, the uncertainty 
in $R_\nu$ is only $\sim 10\%$.  Higher frequencies are desirable
for two additional reasons.  First, since $R_\nu$ is smaller, the source is more
compact, which results in a higher magnification as the source crosses
the caustic, improving the signal-to-noise ratio.  Second, 
Doppler and relativistic effects are larger for radii closer to the black hole.
These effects impose asymmetries in the surface brightness of the disk, 
which further reduce degeneracies between the parameters,
and may even enable a measurement of the black hole spin \citep{ak99}.  

It is also important to note that we have assumed observations of a
single, linear caustic-crossing event.  Observations of multiple
caustic-crossings, or observations of crossings of more complicated
caustic geometries (i.e.\ parabolic folds, cusps, etc.), would likely
remove this degeneracy.  However, the additional complexity of such
geometries may give rise to other complications and new degeneracies;
addressing these issues is beyond the scope of the present paper.

\section{Summary}\label{sec:summary}

Fold caustics are germane to many applications of gravitational
microlensing.  Previous studies have primarily focused on sources with
circular symmetry.  Here we have considered microlensing of elliptical
sources by fold caustics.  We considered only linear fold caustics,
which are generally applicable when the size of the source is much
smaller than the Einstein ring radius of the system, and prove a
useful and analytic approximation to the magnification structure near
real fold caustics.  The total magnification of the two images produced
near such a caustic is proportional to the square-root of the
perpendicular distance to the caustic, and thus diverges as a point
source approaches the caustic.  The chief utility of fold caustics
is that this divergence can be used to achieve high spatial resolution 
and large magnification of faint or otherwise unresolved sources.

By convolving the magnification near a fold caustic with sources of
arbitrary ellipticity, we computed the magnification of a source near
a fold caustic of scale $u_r$ as a function of its ellipticity $e$,
position angle $\phi$ of the major axis with respect to the caustic
normal, area $\pi\rho^2$, and separation from the caustic $\eta\rho$.
We demonstrated that, for $e\ll 1$, the deviation due to ellipticity
is simply $(u_r/\rho)^{1/2}e^2\cos{2\phi}E(\eta)$, where $E(\eta)$ is
simple one-parameter function that can be trivially evaluated
numerically or expressed in terms of elliptic integrals.  The next
higher-order corrections to this expression are of order $e^4$, and
thus it is accurate to $\la 1\%$ for $e\la 0.3$.   From this expression,
it is clear that the deviation due to ellipticity vanishes for $\phi=45^\circ$,
and one can only measure the combination of parameters $e^2\cos{2\phi}$ from
observations of a single fold caustic crossing. For $e\ll1$, the root-mean-square 
light curve deviation due to ellipticity in the range $-1\le \eta \le 1.5$ is 
$0.3e^2\cos{2\phi}$. Surprisingly, we found that the deviation due to 
ellipticity vanishes at some value of $\phi\ge 45^\circ$, for {\it all} values of $e$,
with $\phi>45^\circ$ for $e\ga 0.3$ and increasing with increasing ellipticity.

We considered three applications of microlensing of elliptical sources
near fold caustics.  We first demonstrated that, if most of the
microlensing source stars have rotation periods similar to
chromospherically active, spotted stars in the bulge, then
approximately $\sim 15\%$ should exhibit light curve deviations due to
their oblateness (corresponding to ellipticities $e\ga0.2$) that are
detectable with current observations.  The deviation due to
limb-darkening is qualitatively similar to that due to ellipticity,
such that a uniform elliptical source with parameters $e,\phi$ can
approximately reproduce the light curve of a circular, limb-darkened
source with $\Gamma=-2e^2\cos{2\phi}$.  This may complicate the
interpretation of ultra-precise measurements of limb-darkening with
microlensing.  
We then considered the feasibility of detecting the oblateness of
close-in, giant, planetary companions to the source-stars of
caustic-crossing microlensing events.  We found that planets which are
sufficiently close to provide a detectable reflected light signal are
also generally tidally locked, and therefore rotating too slowly to produce
a substantial deviation due to oblateness.  If tidal locking can be
prevented to preserve an ellipticity of $e\ga 0.2$ even in the close-in
planets, then the oblateness may be observable, especially for
fortuitous geometries (with caustic crossing occurring along the
direction perpendicular to the major or minor axis). Finally, we considered the resolution of
the structure of a quasar accretion disks using microlensing by stars in the
foreground lens of a multiply-imaged quasar.   
We showed that there exists a partial degeneracy between the disk inclination,
scale length, and position angle for the orientation of the projected
ellipse relative to the caustic.  For passbands primarily arising from
the outer portion of the disk, this degeneracy can lead to a factor of
two uncertainty in the determination the disk scale length, which
translates directly into an uncertainty in the black-hole mass, in the
standard thin accretion disk model.  Higher-frequency or
higher-precision observations can reduce this uncertainty.

\acknowledgments 
Work for BSG was supported by a Menzel Fellowship from the Harvard
College Observatory.


\begin{thebibliography}{}


\bibitem[Abe et al.(2003)]{abe03}
Abe, F.\ et al.\ 2003, A\&A Letters, in press (astro-ph/0310410)

\bibitem[Agol \& Krolik(1999)]{ak99} Agol, E.~\& Krolik, J.\ 
1999, \apj, 524, 49 

\bibitem[Albrow et al.(1999)]{albrow99} Albrow, M.~D.~et al.\ 
1999, \apj, 522, 1011 

\bibitem[Albrow et al.(2000)]{albrow00} 
Albrow, M.~D.~et al.\ 2000, \apj, 534, 894 

\bibitem[Albrow et al.(2001)]{albrow01} Albrow, M.~D.~et al.\ 
2001, \apj, 549, 759  

\bibitem[Alcock et al.(2000)]{alcock2000} 
Alcock, C.~et al.\ 2000, \apj, 541, 270 

\bibitem[Ashton \& Lewis(2001)]{al01} Ashton, C.~E.~\& 
Lewis, G.~F.\ 2001, \mnras, 325, 305 

\bibitem[Dominik(2003)]{dominik03}
Dominik, M.\ 2003, MNRAS, submitted (astro-ph/0309581)


\bibitem[Fields et al.(2003)]{fields03} Fields, D.~L.~et al.\ 
2003, \apj, 596, 1305 

\bibitem[Gaudi \& Gould(1999)]{gg99} Gaudi, B.~S.~\& Gould, 
A.\ 1999, \apj, 513, 619 

\bibitem[Gaudi \& Petters(2002)]{gp02} Gaudi, B.~S.~\& 
Petters, A.~O.\ 2002, \apj, 574, 970 


\bibitem[Gaudi, Chang, \& Han(2003)]{gch03} Gaudi, B.~S., 
Chang, H., \& Han, C.\ 2003, \apj, 586, 527 


\bibitem[Goicoechea et~al.(2003)]{gamm03} 
Goicoechea, L.~J., Alcalde, D., 
Mediavilla, E., \& Mu{\~ n}oz, J.~A.\ 2003, \aap, 397, 517 

\bibitem[Gould(2001)]{gould01} 
Gould, A.\ 2001, \pasp, 113, 903 

\bibitem[Gould \& Miralda-Escude(1997)]{gm97} Gould, A.~\& 
Miralda-Escude, J.\ 1997, \apjl, 483, L13 

\bibitem[Graff \& Gaudi(2000)]{gg00} Graff, D.~S.~\& Gaudi, 
B.~S.\ 2000, \apjl, 538, L133 

\bibitem[Grieger, Kayser, \& Refsdal(1988)]{gkr88} Grieger, 
B., Kayser, R., \& Refsdal, S.\ 1988, \aap, 194, 54 

\bibitem[Grieger, Kayser, \& Schramm(1991)]{gks91} Grieger, 
B., Kayser, R., \& Schramm, T.\ 1991, \aap, 252, 508 

\bibitem[Hendry et al.(1998)]{hendry98} Hendry, M.~A., Coleman, 
I.~J., Gray, N., Newsam, A.~M., \& Simmons, J.~F.~L.\ 1998, New Astronomy 
Review, 42, 125 

\bibitem[Hendry, Bryce, \& Valls-Gabaud(2002)]{hendry02} Hendry, 
M.~A., Bryce, H.~M., \& Valls-Gabaud, D.\ 2002, \mnras, 335, 539 

\bibitem[Heyrovsky(2003)]{h03} Heyrovsky, D.\ 2003, \apj, 594, 464

\bibitem[Heyrovsky \& Sasselov(2000)]{hs00} Heyrovsky, D., \& 
Sasselov, D.\ 2000, \apj, 529, 69

\bibitem[Heyrovsky \& Loeb(1997)]{hl97} Heyrovsky, D., \& 
Loeb, A.\ 1997, \apj, 490, 38

\bibitem[Kochanek(2003)]{kochanek03}
Kochanek, C.S.\ 2003, ApJ, submitted (astro-ph/0307422)

\bibitem[Lewis \& Ibata(2000)]{li00} Lewis, G.~F.~\& Ibata, 
R.~A.\ 2000, \apjl, 539, L63 

\bibitem[Jaroszy{\' n}ski \& Mao(2001)]{jm01} Jaroszy{\' 
n}ski, M.~\& Mao, S.\ 2001, \mnras, 325, 1546 

\bibitem[Olech(1996)]{olech96} 
Olech, A.\ 1996, Acta Astronomica, 46, 389 

\bibitem[Seager \& Hui(2002)]{sh02} 
Seager, S.~\& Hui, L.\ 2002, \apj, 574, 1004 

\bibitem[Petters, Levine, \& Wambsganss(2001)]{plw01} 
Petters, A.~O., Levine, H., \& Wambsganss, J.\ 2001, 
Singularity Theory and Gravitational Lensing (Boston : Birkh{\" a}user)  

\bibitem[Schneider, Ehlers \& Falco(1992)]{sef92}
Schneider, P., Ehlers, J., \& Falco, E.\ E.\ 1992, Gravitational  Lenses (Berlin: Springer)

\bibitem[Shakura \& Sunyaev(1973)]{ss73} 
Shakura, N.~I.~\& Sunyaev, R.~A.\ 1973, \aap, 24, 337 

\bibitem[Shalyapin et al.(2002)]{shalyapin02} Shalyapin, V.~N., 
Goicoechea, L.~J., Alcalde, D., Mediavilla, E., Mu{\~ n}oz, J.~A., \& 
Gil-Merino, R.\ 2002, \apj, 579, 127 

\bibitem[Udalski et al.(2000)]{udalski2000} Udalski, A., Zebrun, 
K., Szymanski, M., Kubiak, M., Pietrzynski, G., Soszynski, I., \& Wozniak, 
P.\ 2000, Acta Astronomica, 50, 1 

\bibitem[Valls-Gabaud(1998)]{vallsgabaud98} Valls-Gabaud, D.\ 1998, 
\mnras, 294, 747 

\bibitem[van Belle(1999)]{vb99} 
van Belle, G.~T.\ 1999, \pasp, 111, 1515 

\bibitem[Wo{\' z}niak et al.(2000)]{wozniak00} 
Wo{\' z}niak, 
P.~R., Udalski, A., Szyma{\' n}ski, M., Kubiak, M., Pietrzy{\' n}ski, G., 
Soszy{\' n}ski, I., {\. Z}ebru{\' n}, K.\ 2000, \apjl, 540, L65 

\bibitem[Wyithe, Webster, \& Turner(2000)]{wyithe00} 
Wyithe, J.~S.~B., Webster, R.~L., \& Turner, E.~L.\ 2000, \mnras, 318, 1120 

\bibitem[Yonehara(2001)]{yonehara01} 
Yonehara, A.\ 2001, \apjl, 548, L127 

\end{thebibliography}
\end{document}